\documentclass[preprint,authoryear,12pt]{elsarticle}

\usepackage{graphicx}
\usepackage{epstopdf}
\usepackage[dvipdfm]{geometry}

\usepackage{longtable}
\usepackage{multirow}
\usepackage{array}
\usepackage{amsthm,amsmath}
\usepackage{epsfig}
\usepackage{amssymb}
\usepackage{amsthm,amsmath}
\usepackage{threeparttable}
\usepackage[justification=centering]{caption}
\usepackage{url}
\usepackage{color}
\usepackage[figuresright]{rotating}
\usepackage{tabularx}
\usepackage{array,booktabs}
 \usepackage{setspace}
\usepackage{epsfig}
\usepackage{amssymb}
\usepackage{amsthm,amsmath}
\usepackage{threeparttable}
\usepackage[justification=centering]{caption}
\usepackage{url}
\graphicspath{{figures/}}
\journal{arXiv}

\begin{document}

\begin{frontmatter}


\title{Oil price shocks, road transport pollution emissions and residents' health losses in China}

\author{Sheng YANG}
\ead{yangsheng11111@163.com}
\author{Ling-Yun HE \corref{cor1}}
\ead{lyhe75@gmail.com}
\address{College of Economics and Management, China Agricultural University, Beijing 100083, China}

\cortext[cor1]{Prof. Dr. HE L.-Y. is the corresponding author. Dr. HE is a full professor of energy economics and environmental policies. YANG is a Ph.D. candidate supervised by Dr. HE.  The authors contribute equally in the project. HE conceived the whole project, proposed the hypothesis and designed the scenarios. YANG calculated the results based on the scenarios under Dr. HE's supervision. HE and YANG co-wrote the manuscript. This project is supported by the National Natural Science Foundation of China (Grant Nos. 71273261 and 71573258).}

\begin{abstract}
In recent years, China's rapid economic growth resulted in serious air pollution, which caused substantial losses to economic development and residents' health. In particular, the road transport sector has been blamed to be one of the major emitters. During the past decades, fluctuation in the international oil prices has imposed significant impacts on the China's road transport sector. Therefore, inspired by \cite{Li2005}, we propose an assumption that China's provincial economies are independent ``economic entities''. Based on this assumption, we investigate the China's road transport fuel (i.e., gasoline and diesel) demand system by using the panel data of all 31 Chinese provinces except Hong Kong, Macau and Taiwan. To connect the fuel demand system and the air pollution emissions, we propose the concept of pollution emissions elasticities to estimate the air pollution emissions from the road transport sector, and residents' health losses by a simplified approach consisting of air pollution concentrations and health loss assessment models under different scenarios based on real-world oil price fluctuations. Our framework, to the best of our knowledge, is the first attempt to address the transmission mechanism between the fuel demand system in road transport sector and residents' health losses in the transitional China.
\end{abstract}


\end{frontmatter}

\section{Introduction}
\label{intro}

Since the year of 1978 when Chinese central government began to promote its momentous economic reforms, China has been keeping a miraculously high speed of economic growth, with annual growth rates averaging 10\% over the past three decades. This former centrally planned economy successfully transformed itself into the socialist market economy as we know it, and by now becomes the world's second largest economy by nominal GDP, and (arguably) the world's largest by purchasing power parity (PPP) according to the World Economic Outlook (October 2014 Edition) by the International Monetary Fund (IMF) \footnote{For more details, please see: \url{http://www.imf.org/external/pubs/ft/weo/2014/02/weodata/index.aspx}.}.

However, at the same time, China's rapid economic growth also resulted in serious environmental issues, especially on the air pollution emissions, causing substantial losses to economic development and residents' health \citep{Hao2005, Brajer2006, Hammitt2006, He2013, Ouyang2013}. It is estimated that about two-thirds of China's cities have not attained the ambient air quality standards applicable to urban residential areas \citep{Hao2005}. According to \cite{WorldBank2007}, the economic burden of premature mortality and morbidity associated with air pollution in China accounted for 157.3 billion RMB yuan in 2003, or 1.16\% of GDP. \cite{chen2013} estimate that total suspended particles (TSPs), one of the major air pollution emissions, is causing the 500 million residents in Northern China to lose more than 2.5 billion life years of life expectancy.

China's road transport sector, which is experiencing rapid development\footnote{For example, in recent years, the China's road transport sector has enormous growth in the road transport infrastructure investment, etc. For more information, please refer to the Appendix section.}, is found to be one of the major emitters, and responsible for serious air pollution and huge residents' health losses \citep{He2013, Chen2014}, especially in urban areas \citep{Hao2005, Guo2010}. In 2012, the total vehicle emissions in China reached to 46.12 million tons. Specifically, the emissions of $\mathrm{NO}_{x}$, $\mathrm{PM}_{2.5}$, $\mathrm{HC}$, and $\mathrm{CO}$ were 6.4, 0.622, 4.382 and 34.71 million tons, respectively, from road transport sector\footnote{Data source: China Vehicle Emission Control Annual Report 2013 by the Ministry of Environmental Protection of China. http://www.zhb.gov.cn/gkml/hbb/qt/201401/t20140126\_266973.htm}. \cite{Hao2005} find that 74\% of the ground $\mathrm{NO}_{x}$ in Beijing were attributed to vehicles while only 2\% and 13\%  were emitted by power plants and industry respectively. They also find that vehicle exhaust accounted for 46\% of the total VOCs emission in Beijing. \cite{Guo2010} estimate the total economic costs of health impacts due to air pollution contributed from transport in Beijing during 2004 -- 2008 were 272, 297, 310, 323, 298 million US dollars (mean values) respectively, which accounted for 0.52\%, 0.57\%, 0.60\%, 0.62\%, and 0.58\% of annual local GDP.

As we all know, gasoline and diesel price changes have some significant impacts on the fuel demand from the road transport sector. After the 2008 financial crisis, crude oil prices experienced drastic fluctuations. For example, from 2008 to 2009, the annual average crude oil prices of WTI and Brent fell by 38\% and 36\%, respectively. In 2010, crude oil prices of WTI and Brent increased 28\% and 29\% than those in 2009, respectively. However, WTI crude oil price in 2013 was only 4\% higher than that in 2012; in contrast, Brent crude oil price decreased by 3\%. From August 2014, the international crude oil price continued to fall from \$100 per barrel to around \$50 per barrel in February 2015, nearly 50 percent decline\footnote{Data source: U.S. Energy Information Administration (EIA), http://www.eia.gov.}. As a result, in China, the retail prices of gasoline and diesel also appeared rare ``thirteen losing streak''. Therefore, there are some extremely important issues: What do oil price changes influence the fuel demand from the road transport sector in China? How does oil price change transmit through the fuel demand to the road transport pollution emissions and residents' health losses? Unfortunately, by now, the current studies in existing literature failed to address these important questions.

Furthermore, after the fiscal decentralization reforms in China, the role of local governments in China is the key to understand the rapid development of road transport sector and residents' health losses caused by environmental pollution\footnote{Generally speaking, local governments have both the willingness and financial capacity to promote the development of the road transport sector, consequently making the road transport sector one of the major emitters causing huge health losses; furthermore, to cope with serious pollution and health issues, local governments began to implement traffic policies to alleviate road transport pollution. In the appendix section of this article, we will elaborate on the fiscal decentralization and road transport growth in China}. Therefore, inspired by \cite{Blanchard2001} and \cite{Li2005}, we propose a key assumption that China's provincial economies are independent ``economic entities'' acting as the equivalent of a business company in this broad context of the transitional China. In addition, please also note that these provincial ``economic entities'' are price-takers rather than price-makers in terms of oil prices, which are mainly determined by global markets.

Therefore, by using the road transport panel data from all 31 Chinese provincial level distracts except Hong Kong, Macau and Taiwan, we estimate the road transport fuel (i.e. gasoline and diesel) demand system through an Almost Ideal Demand System (AIDS) model. Finally, by means of pollution emissions elasticities as well as air quality and health evaluation models, we investigate the impact of pollution emissions and residents' health losses from road transport under different scenarios based on real world oil price fluctuations. Our framework, to the best of our knowledge, is the first attempt to study the underlying transmission mechanisms of the fuel demand and residents' health losses in China in the context of dramatically changing oil prices.

The remaining of this paper comprises four main sections. The Section 2 models the framework of this study, which mainly consists of the AIDS model, air pollution emissions elasticities, air pollution concentrations model, and health loss assessment model, etc. Applying these methods, the data and some key parameters are posited in the Section 3. In the Section 4, we present the empirical results of fuel demand elasticities, pollution emissions elasticities, air pollution emissions, and health losses. The Section 5 concludes with directions for further research.  Finally, if interested, our reader(s) may find the Appendix very informative to read, which provides many interesting facts, and relatively independent literature review, and our justification of the hypothesis. Although this section is somehow separable and stand-alone from our main story, one may better understand not only the hypothesis we applied in the study, but also (hopefully) Chinese economy from a new perspective.

\section{Methodology}

\subsection{The price and expenditure elasticities of gasoline and diesel demand}

The first step of this study is to estimate the price and expenditure elasticities of fuel (gasoline or diesel) demand for all vehicle types by means of the Almost Ideal Demand System (AIDS) model developed by \cite{deaton1980}. In this paper, one of the basic assumptions is that the  vehicles' fuel (gasoline or diesel) demands are weakly separable from other vehicle types and other economic sectors' fuel demands in the provincial economic entities' budgets. Provided this assumption, the AIDS model is defined as follows:

\begin{equation}\label{AIDS}
w_{i}=\alpha_{i}+\sum ^{N} _{j=1}\gamma_{ij}\ln p_{j}+\beta_{i}\ln (\frac{X}{P})+\nu_{i}
\end{equation}

\begin{equation}\label{lnP}
\ln P=\alpha_{0}+\sum ^{N} _{j=1} \alpha_{j}\ln p_{j}+\frac{1}{2}\sum ^{N} _{i=1}\sum ^{N} _{j=1}\gamma_{ij} \ln p_{i} \ln p_{j}
\end{equation}
where both $i$ and $j$ indicate vehicle types. When vehicle type $i$ (or $j$) is determined, the fuel type it uses is also determined. $N$ stands for the number of vehicle types ($N$=10). $w_{i}$ is the vehicle fuel expenditure share of vehicle type $i$. $p_{i}$ (or $p_{j}$) means fuel price of $i$th (or $j$th) type of vehicle. If $p_{i}$ = $p_{j}$, the $i$th and the $j$th vehicle types employ the same kind of fuel; otherwise ($p_{i}$ $\neq$ $p_{j}$), they use different types of fuels\footnote{This study includes a total of ten vehicle types using gasoline or diesel as fuel. For more information in details, please see the Section 3}. $X$ represents provincial expenditure on fuel in the demand system. $P$ indicates the translog price index. $\gamma_{ij}$ and $\beta_{i}$ are parameters to be estimated. $\nu_{i}$ is the error term.

Specifically, the fuel consumption of vehicle type $i$ (or $j$) is estimated by using these data including the vehicle population, vehicle mileage travelled (VMT) and fuel (gasoline or diesel) economy. Then, with the fuel (gasoline or diesel) price, fuel expenditure of vehicle type $i$ (or $j$) is calculated. By this mean, the provincial expenditure $X$ and fuel expenditure share of vehicle type $i$ (i.e., $w_{i}$) can also be estimated.

The restrictions from neoclassical demand theory can be imposed on the parameters of the model (\ref{AIDS}), which include:
\begin{itemize}
	\item the Slutsky symmetry condition given by $\gamma_{ij}=\gamma_{ji}$ for any $i$ and $j$;
	\item the homogeneity condition imposed as $\sum ^{N} _{j=1} \gamma_{ij}=0$ for any $j$;
	\item the adding-up restriction  given by $\sum ^{N} _{i=1} \alpha_{i}=1$, $\sum ^{N} _{i=1} \beta_{i}=0$ and $\sum ^{N} _{i=1} \gamma_{ij}=0$.
\end{itemize}

The price elasticity of fuel demand is defined as:
\begin{equation}\label{price-ela}
e_{ij}=-\delta_{ij}+w_{i}^{-1}(\gamma_{ij}-\beta_{i}(\alpha_{j}+\sum^{N}_{i=1}\gamma_{ij}\ln p_{i}))
\end{equation}
where $\delta_{ij}$ is the Kronecker delta (1 if $i=j$, and 0 otherwise).
The fuel expenditure elasticity is given by:
\begin{equation}\label{exp-ela}
e_{i}=1+\frac{\beta_{i}}{w_{i}}
\end{equation}

\subsection{The air pollution emissions elasticities}

The second step is to investigate the fuel demand changes arising from changes of air pollutants. Inspired by the concept of nutrient elasticities applied in some studies on the human nutritional intake \citep{huang1996, zheng2012}, we propose the concept of air pollution emissions elasticities (see Eqs.(5) and (6)), which reflect how the changes in different types of vehicle activities impact on the air pollution emissions. Specifically, the air pollution emissions refer to the emissions of three major pollutants from various vehicle types, including $\mathrm{CO}$, $\mathrm{NO}_{x}$, and $\mathrm{PM}_{2.5}$. Based on the demand elasticities defined above, the air pollution emissions elasticities are given by:
\begin{equation}\label{pi}
    \pi_{ki}=\sum_{i=1}^{N}e_{ij}a_{ki}q_{i}/\psi_{k}
\end{equation}
\begin{equation}\label{eta}
    \eta_{k}=\sum_{i=1}^{N}e_{i}a_{ki}q_{i}/\psi_{k}
\end{equation}
\begin{equation}\label{psi}
    \psi_{k}=\sum_{i=1}^{N}a_{ki}q_{i}
\end{equation}
where $\pi_{ki}$ denotes the price elasticity of air pollution emissions, which illustrates the weighted average of all own- and across-price emissions elasticities for air pollutant $k$ in response to the price of fuel for the $i$th vehicle type; $a_{ki}$ is the pollution emission factor, which means that the pollutant $k$ emission from unit mileage of $i$th vehicle type; $\eta_{k}$ represents the expenditure elasticity of air pollutant emissions, which indicates the weighted average of all air pollutant $k$ emissions elasticities with respect to expenditures; $q_{i}$ denotes the quantity of fuel used by vehicle type $i$; $\psi_{k}$ stands for the total amount of pollutant emissions from all types of vehicles.

\subsection{A simplified air-quality-induced-health-losses estimation}

The third step is to estimate the changes in air pollutant concentrations caused by road transport emissions, and then to evaluate the induced health damages.

In this study, we use the air concentration model to estimate the changes in air quality. To simplify the question, we employ a fixed box model\footnote{The issue of the estimation of air pollutant concentration is very complicated, especially on account of the geographical and environmental heterogeneities among different provinces in China. Due to the availability of the relevant data across the provinces over time in China, this paper uses the box model to simplify the problem, although the box model has obvious limitations in this context, e.g. this model simulates the formation of pollutants within the box, providing no information on the local concentrations of the pollutants. For this reason, the box model may have obvious limitation in modeling the particle concentrations within a localized environment, where concentrations and thus particle dynamics are highly influenced by local changes to the wind field and emissions \citep{Holmes2006}.} whereby China is represented by a parallelepiped with the uniform pollutant dispersion according to \cite{Chen2014}.
\begin{equation}\label{airq}
C_{k}=b_{k}+\frac{S_{k} \cdot L}{H \cdot u}
\end{equation}
where $C_{k}$ stands for the concentration of pollutant $k$ in the entire site ($ \mu g/ m^{3}$); $b_{k}$ the background concentration of pollutant $k$ under natural conditions ($\mu g/m^{3}$); $S_{k}$ the emission rate of pollutant $k$ ($\mu g/s/m^{2}$); $L$ the length (m); $H$ the mixing height (m); $u$ the wind velocity (m/s).

Given the difficulty of obtaining the meteorological parameters ($S_{k}$, $L$, $H$, $u$), specifically in China at the national level, we use the simplified model \ref{qtq1}. In this model, we assume these parameters ($b_{k}$, $L$, $H$, $u$) are constant under the baseline situation ($S_{k1}$, $C_{k1}$) and the future situation ($S_{k2}$, $C_{k2}$). By this simplified model, we can estimate the change in the concentration of the air pollutant $k$.

\begin{equation}
\label{qtq1}
   \frac{S_{k2}}{S_{k1}}=\frac{C_{k2}-b_{k}}{C_{k1}-b_{k}}=\frac{E_{k2}}{E_{k1}}
   \end{equation}
where $S_{k1}$ and $S_{k2}$ represent the baseline emission rate and the future emission rate of pollutant $k$, respectively. The $C_{k1}$ and $C_{k2}$ stand for respectively the annual average baseline and the future concentration levels.

We further assume that $E_{k1}$ and $E_{k2}$ are the baseline emissions and future emissions of pollutant $k$, respectively, from road transport sector in China. Then the ratio of $E_{k2}$ to $E_{k1}$ reflects the change in the pollutant $k$ emissions from road transport sector. Under the assumption of the uniform emission rate, it is reasonable to think of the ratio of $S_{k2}$ to $S_{k1}$ as the ratio of the $E_{k2}$ to $E_{k1}$. Considering data reliability, we take the Chinese vehicle pollution emissions data in 2012 as baseline emissions $E_{k1}$\footnote{Data source: China Vehicle Emission Control Annual Report 2013, which is released by the Ministry of Environmental Protection of China. http://www.zhb.gov.cn/gkml/hbb/qt/201401/t20140126\_266973.htm} (see Table \ref{tab:0}).

\begin{center}
\resizebox{1\textwidth}{!}{
\begin{threeparttable}
\small
  \caption{Summary of parameters of emissions, air concentrations, and Exposure-Response (ER) coefficients\tnote{a}}
  \centering
 \label{tab:0}

 \begin{tabular}{ccccc}

 \hline

  Air pollutants  & Background concentrations\tnote{b}   &  ER coefficients\tnote{c}   & Baseline concentration\tnote{d} & Baseline emissions\tnote{e}\\
  \hline
  $\mathrm{CO}$        &    1                          &  3.7                                & 1.3                      &     34.71  \\
  $\mathrm{NO}_{x}$    &    10                         &  0.13                               & 47                       &      6.4  \\
  $\mathrm{PM}_{2.5}$  &    39                         &  0.042                              & 44.7                     &      0.404   \\

  \hline

 \end{tabular}
 \begin{tablenotes}
 \item[a] We assume that all of the parameters are China's national level value in 2012 due to limited data availability in China.
 \item[b] For the background concentration of $\mathrm{CO}$, please see \cite{Li2011}. The background concentrations of $\mathrm{NO}_{x}$ and $\mathrm{PM}_{2.5}$ see \cite{Wu2010} and \cite{Chen2014}, respectively. Unit: $mg/m^{3}$ ($\mathrm{CO}$), $\mu g/m^{3}$ ($\mathrm{NO}_{x}$, $\mathrm{PM}_{2.5}$).
 \item[c] The ER coefficients for acute exposure are expressed in mortality percentage change per $\mu g/m^{3}$ ($mg/m^{3}$) change of pollutant concentration.
 For the ER coefficients of $\mathrm{CO}$ and $\mathrm{NO}_{x} $, please see \cite{Shang2013}; and for the $\mathrm{PM}_{2.5}$, please see \cite{Chen2014}.
  \item[d] For the baseline concentration of $\mathrm{CO}$ and $\mathrm{NO}_{x} $, please see \cite{Shang2013}. The baseline concentration of $\mathrm{PM}_{2.5}$,  please see \cite{Chen2014}. Unit: $mg/m^{3}$ ($\mathrm{CO}$), $\mu g/m^{3}$ ($\mathrm{NO}_{x}$, $\mathrm{PM}_{2.5}$).
 \item[e] In 2012, the $\mathrm{PM}_{10}$ emissions from road transportation in China were 0.622 million tons (Data source: China Vehicle Emission Control Annual Report 2013.). We use $0.65$ as the $\mathrm{PM}_{10}$--$\mathrm{PM}_{2.5}$ conversion factor \citep{Chen2014}. Therefore, the $\mathrm{PM}_{2.5}$ emissions from road transportation were 0.404 million tons. Unit: million tons.
\end{tablenotes}
\end{threeparttable}
}
\end{center}

Once the change in air pollutant concentration level is determined, the consequent risks in health outcomes can be assessed using exposure-response (ER) functions from epidemiological studies which typically include mortality (acute and chronic), respiratory hospital admission, cardiovascular hospital admission, restricted activity day, work loss day, asthma attack, etc \citep{Chen2014}. Due to the data availability, the quantifiable health effects in our analysis only include the acute mortality\footnote{As a consequence, please note that the real health losses might be significantly underestimated in this study. Even so, the estimated results are formidably substantial (please see  Table \ref{tab:5}).}. The number of premature deaths from acute exposure \citep{Chen2014} is estimated by
 \begin{equation}\label{AM}
  Case^{AM}= \sum _{k} ER^{AM}_{k} \times C_{k} \times N \times M
  \end{equation}
where $ER^{AM}_{k}$ represents the ER coefficients for mortality from acute exposure to the pollutant $k$. $C_{k}$ denotes the concentration level of pollutant $k$ (see Eq.\ref{qtq1}). $N$ illustrates the total population exposed to the air pollution. $M$ means the overall mortality rate.

In addition, in order to avoid the limitations of linear health effects model, this paper also employs the non-linear health effect models \citep{Ostro2004,Corbett2007,Winebrake2009} to evaluate the acute mortality (please see Eqs.(\ref{AMrr}), (\ref{AF}) and (\ref{AMe})). The results from the linear and non-linear models will be compared to get more robust and accurate estimations.
%(Lepeule2012)

\begin{equation}\label{AMrr}
  RR_{k}=e^{\beta_{k}(C_{k}-b_{k})}
  \end{equation}
  \begin{equation}\label{AF}
   AF_{k}=\frac{RR_{k}-1}{RR_{k}}
     \end{equation}
   \begin{equation}\label{AMe}
   E_{k}= AF_{k} \times N \times M
   \end{equation}
where $RR_{k}$ represents the relative risk of pollutant $k$ (i.e. $\mathrm{CO}$, $\mathrm{NO}_{x}$ and $\mathrm{PM}_{2.5}$). $\beta_{k}$ stands for the ER coefficients for acute exposure of pollutant $k$ (see the Table \ref{tab:0}); $C_{k}$ is the pollutant concentration for the case under study ($mg/m^{3}$ for $\mathrm{CO}$; $\mu g/m^{3}$ for $\mathrm{NO}_{x}$ and $\mathrm{PM}_{2.5}$); and $b_{k}$ the background concentration of pollutant $k$ under natural conditions ($mg/m^{3}$ for $\mathrm{CO}$; $\mu g/m^{3}$ for $\mathrm{NO}_{x}$ and $\mathrm{PM}_{2.5}$); $AF_{k}$ means the attributable fraction of deaths to air pollutant $k$; $E_{k}$ is the total number of cases of premature mortality from acute exposure to the pollutant $k$.

For the residents' health loss assessment, we apply the value of statistical life (VOSL) method, which represents an individual's willingness to pay (WTP) for a marginal reduction in the risk of death \citep{Zhang2008,Kan2004}:
\begin{equation}
\label{VOSL}
VOSL=VOSL_{BL} \times \left(\frac{I}{I_{BL}}\right)^{e}
\end{equation}
where $VOSL$ and $VOSL_{BL}$ are the VOSL of national level and baseline level, respectively. While $I$ and $I_{BL}$ represent the income per capita of the national and baseline levels, respectively. $e$ is the elastic coefficient of WTP and assumed to be 1. Considering data availability, the baseline $VOSL_{BL}$ is obtained from \cite{Zhang2008}, which was \$ 79,839 in Chongqing, one of the four province-level municipalities in China, in 2004.

\section{Data and key parameters description}

Firstly, the vehicle population data are specifically divided into ten vehicle categories\footnote{This paper uses the classification system of the Ministry of Public Security, and detailed vehicle category data from the official statistical reporting by the National Bureau of Statistics of China (NBSC) \citep{He2013,Kayser2000}.} (please see Table \ref{tab:4}). The panel data set covers approximately 3100 data in 31 provinces in mainland China from 2002 to 2011\footnote{Data source: the NBSC, http://data.stats.gov.cn/index.}.

Secondly, the annual average vehicle mileage traveled (VMT) data in China is used \footnote{Some previous studies and estimates on China's VMT data can be seen in \cite{He2005,Huo2012a}}, which is closely related to the economic growth level and road traffic infrastructure \citep{Huo2012a}\footnote{The VMT is also affected by vehicle age and other factors \citep{He2005,Huo2012a}. Due to data availability, these factors are not considered in this paper.}. As a matter of fact, there exist huge differences between different regions of China in geographic conditions, economic development and transportation infrastructure, etc. However, the existing China's VMT estimates in different provinces are not readily available or substantially insufficient. Therefore, we use the national average VMT data in 2002 as the baseline data. Then, by using the 31 provincial average passenger and freight transport distances, we estimate the provincial annual average VMT data for each type of vehicles from 2002 to 2011. Our VMT estimates are supported by many survey papers and empirical studies in existing literatures (please see Table \ref{tab:4}), indicating that our estimated results are reasonably acceptable.

%\begin{sidewaystable}
%\begin{center}
%\small
\begin{threeparttable}

  \caption{The annual average VMT, fuel economy and pollution emission factors\tnote{a}}
  \centering
 \label{tab:4}
 \begin{tabular}{cccccc}

 \hline

\multicolumn{1}{c}{\multirow{2}{*}{Vehicle Types\tnote{b}}} & Baseline VMT\tnote{c}  & Fuel Economy\tnote{d} & \multicolumn{3}{c}{Pollution Emission Factors (g/km)}             \\
 \cline{4-6}
       &    (1000 km)          &  (L/100 km)     &   $\mathrm{CO}$\tnote{e} &  $\mathrm{NO}_{x}$\tnote{f} & $\mathrm{PM}_{2.5}$\tnote{g} \\
  \hline
   LPV-D  & 48.6   &  32.6   & 6.7   & 12.772  & 0.2567\\
   MPV-G  & 47.3   &  25.97  & 4.1   & 0.47  & 0.126 \\
   SPV-G  & 33.6   &  9     & 1.57   & 0.37  & 0.117   \\
   MNPV-G & 34     &  6.38  & 3.33   & 1.24  & 0.09  \\
   HDT-D  & 50     &  24.9  & 6.3    & 10.2  & 0.23  \\
   MDT-D  & 24     &  15    & 1.5    & 6.4   & 0.11  \\
   LDT-D  & 20     &  12.9  & 2.9    & 3.2   & 0.17  \\
   MNT-G  & 38.4   &  7.96  & 1.57   & 0.37  & 0.09 \\
   PB-D   & 57.2   &  33    & 6.7    & 12.772  & 0.35 \\
   Taxi-G & 74.9   &  8.7   & 0.927  & 0.148   & 0.117\\

  \hline

 \end{tabular}
\footnotesize
 \begin{tablenotes}
   \item[a]  Although the fuel economy and pollution emission factors may change over time and space in real world problems, we have to make some simplifications due to data availability in China. Particularly, from a nationwide perspective, the fuel economy is likely to be getting better over time by virtue of technological progress, policies, etc. As a consequence, this paper may over-estimate the fuel consumption relative to the actual fuel consumption of road transport sector in China. Nevertheless, it is difficult to determine whether the fuel elasticities are over- or under-estimate) due to the complexity of the real world problems in China.
  \item[b] LPV: Large passenger vehicles; MPV: Medium passenger vehicles; SPV: Small passenger vehicles; MNPV: Mini passenger vehicles; HDT: Heavy duty trucks; MDT: Medium duty trucks; LDT: Light duty trucks; MNT: Mini trucks; PB: Public buses. D: diesel; G: gasoline.
  \item[c] Baseline VMT refer to the 2002 national average VMT. The VMT of LPV-D, MPV-G, SPV-G, MNPV-G, HDT-D, MDT-D, LDT-D and MNT-G, see China Energy Databook v.7.0, October 2008. The VMT of PB-D and Taxi-G see \cite{Huo2012a}.
   \item[d] The fuel economy: LPV-D (\cite{Zhang2014}); MPV-G, SPV-G, HDT-D, MDT-D, LDT-D,and MNT-G (\cite{Huo2012b}); MNPV-G (\cite{Wang2005}); PB-D (\cite{Shen2014}); Taxi-G (\cite{Hu2012}).
   \item[e] The pollution emission factors of $\mathrm{CO}$: LPV-D, PB-D (\cite{Wang2011b}); MPV-G (\cite{Huo2012c}), SPV-G, MNPV-G, and MNT-G (\cite{Wang2005}); HDT-D, MDT-D, and LDT-D (\cite{Huo2012d}); Taxi-G (\cite{Hu2012}).
 \item[f] The pollution emission factors of $\mathrm{NO}_{x}$: LPV-D, PB-D (\cite{Wang2011b}); MPV-G (\cite{Huo2012c}); SPV-G, MNPV-G, and MNT-G (\cite{Wang2005}); HDT-D, MDT-D, and LDT-D (\cite{Huo2012d}); Taxi-G (\cite{Hu2012}).
 \item[g] The pollution emission factors of $\mathrm{PM}_{2.5}$: LPV-D (\cite{Cheng2010}); MPV-G (\cite{Yan2010}); SPV-G (\cite{He2005}); HDT-D, MDT-D, and LDT-D (\cite{Huo2012d}); PB-D (\cite{Oanh2010}); MNT-G (\cite{Yao2011}). Due to the unavailable emission factors of $\mathrm{PM}_{2.5}$ for Taxi-G and MNPV-G, we assume that they are the same as SPV-G and MNT-G
\end{tablenotes}
\end{threeparttable}
\vspace{10 pt}

Thirdly, in 2012, the number of China's total population was 1,354,040,000, while the overall mortality rate was 0.715\%. According to \cite{Zhang2008}, the Chinese national average VOSL, which was calculated after inflation and exchange rate adjustments, was 855,642.81 RMB yuan in 2012\footnote{For the inflation rate and exchange rate of RMB to USD, see the NBSC, http://data.stats.gov.cn/index}.

Finally, the China's provincial gasoline and diesel annual average retail prices, from 2002 to 2011, are used\footnote{Data source: The Wind Information Co., Ltd. (Wind Info), http://www.wind.com.cn/}. In China, gasoline and diesel retail prices are mainly determined by international oil prices\footnote{Data source: Oil price management (tentative) of China introduced by the National Development and Reform Commission(NDRC), http://bgt.ndrc.gov.cn/zcfb/200905/W020120820331590086949.pdf. Specifically, in China, the National Development and Reform Commission (NDRC) sets the guiding retail prices of gasoline and diesel for the provinces (including the provincial level autonomous regions and municipalities) or the central cities. While, the guiding retail prices are mainly determined by international oil prices (as pricing benchmarks), and then adjusted (slightly) according to some domestic factors such as average manufacturing costs, taxes, circulating costs, profits, and so on.}. The international oil prices are the pricing benchmarks of China's domestic oil prices, and fundamentally important in China's gasoline and diesel pricing mechanisms. In this regard, the road transport sector, however, has no say in the fuel pricing, and is merely price taker rather than price maker\footnote{According to China's oil pricing mechanism mentioned above, the road transport sector at the provincial level has no say at all in oil (or fuel) pricing as a matter a fact. In current literature, empirical studies \citep{Mu2011,Chen2009} also show that China has no pricing power over international oil prices. As a consequence, the road transport sector in China is merely price taker rather than price maker.}. In this paper, we apply the China's provincial gasoline and diesel annual average retail prices.

\section{Empirical estimation}

\subsection{Fuel demand elasticities}

In the existing literature, many studies investigate empirically on the price elasticity of fuel (please see Table \ref{tab:5} for more details). However, these results vary significantly in terms of different time periods, areas (e.g., different countries/regions and economic sectors), data and methods (e.g., time series or cross section data, regression or meta-analysis methods), etc. Furthermore, estimates vary greatly both between and within geographical areas of study for long- and short-run price elasticities. Typically, \cite{Graham2002} find that short-run price elasticity estimates range from $-0.34$ to $-0.5$ within the OECD countries, while long-run price elasticities even range from $-0.23$ to $-1.35$. Even for fuel price elasticities in China, huge differences can be found in estimates among current studies (see Table \ref{tab:5}).

In this study, therefore, the first step is to estimate and clarify the price elasticities of fuel demand (own-price and expenditure elasticities) for different vehicle types, which indicate their fuel demand changes caused by the changes in fuel prices. All fuel demand elasticities are estimated on the basis of parameter estimates and sample means of explanatory variables. The standard errors of these elasticities are approximated using the delta method. The fuel demand elasticities contain 110 estimates of own- and cross-price as well as expenditure elasticities for ten vehicle categories across the entire sample\footnote{Due to the limited space, the cross-price elasticities estimates are not listed in this paper. The full results of cross-price elasticities are available upon request.}. The own-price and expenditure elasticities are reported in Table \ref{tab:6}.

 Our results estimated by AIDS model in Table \ref{tab:6} demonstrate that all own-price elasticities are negative and statistically significant, which is consistent with the most previous studies (see Table \ref{tab:5}). Expenditure elasticities are positive and statistically significant. The estimated own-price elasticities vary across different vehicle categories. The own-price elasticities values (in absolute values) are less than 1 for SPV-G ($-0.552$), MNPV-G ($-0.629$), HDT-D ($-0.823$), MDT-D ($-0.507$),  LDT-D ($-0.799$),  MNT-G ($-0.657$),and PB-D ($-0.442$), implying that these are not sensitive to fuel price changes. Especially, for PB-D, the value is just $-0.442$. A plausible explanation is that local governments usually support their urban transit systems with price subsidies (e.g. Beijing prolonged transit subsidies and low fares). Therefore, changes in the diesel price have limited impacts on public bus system.

Conversely, fuel (gasoline or diesel) demands for LPV-D, MPV-G and Taxi-G are price-sensitive with their own-price elasticities values (in absolute values) greater than 1. In particular, the price elasticity for Taxi-G is $-1.263$, which is much larger than those reported in many previous studies (please see Table \ref{tab:5}). This may be explained plausibly by the fact that most of existing studies estimate gasoline (or diesel) demand elasticity at the national/country level, therefore other sectors such as industrial sector, agricultural sector, etc., are mixed together in those studies.

It is noteworthy that the different model selections may also cause the significant difference in the elasticities estimated in current literature. Therefore, to avoid the (possible) bias caused by the single model selection, we employ also double-log regression model \citep{Graham2002,Chang2014,Ramanathan1999,Cheung2004} on fuel demand elasticities, and compare the estimates with the results obtained from the AIDS model. Table \label{tab:6} shows that the results obtained from the double-log model are qualitatively consistent with those from the AIDS model. Generally speaking, fuel demand elasticities estimated by the two different models are relatively large. These results from the double-log regression and AIDS models indicate that our estimates are consistent and reliable across different models.

\begin{center}
\resizebox{1\textwidth}{!}{
\begin{threeparttable}
\small
  \caption{Price elasticities of gasoline demand in previous studies}
  \label{tab:5}
   \centering
     \begin{tabular}{lll}
    \hline
    Study & Research methods & Price elasticity value \\
    \hline
    \cite{Dahl1991}& Econometric modeling analysis &  Short-run:$-0.26$; Long-run:$-0.86$     \\
    \cite{Espey1998} & Meta-analysis of literature & Short-run:$-0.26$; Long-run:$-0.58$ \\
    \cite{Kayser2000}& Regressions & Households in USA, Short-run:$-0.23$    \\
    \cite{Graham2002}& Literature survey & OECD countries: Short-run:$(-0.34, -0.5)$; \\
                                                       &     & \quad  \quad  \quad  \quad  \quad  \quad  \quad  \quad Long-run:$(-0.23, -1.35)$\tnote{a}    \\
     \cite{West2004}& AIDS & Households in USA: $(-0.18, -0.74)$    \\
     \cite{Cheung2004}& Regressions & China: Short-run:$-0.19$; Long-run:$-0.56$    \\
    %  \cite{Brons2008} &  & Short-run:$-0.34$; Long-run:$-0.84$ \\
     \cite{Lin2011}  & Regressions & Transport sector in China: $-0.269$    \\ %\tnote{b}
    %  \cite{Lin2013} &  Regression & Intermediate-run in China:$-0.196 \sim -0.497$ \tnote{a}     \\
    \hline
    \end{tabular}
     \begin{tablenotes}
   \item The $(-0.34, -0.5)$ show that the short-run price elasticity estimates range from $-0.34$ to $-0.5$. The $(-0.23, -1.35)$ show that the long-run price elasticity estimates range from -0.23 to -1.35.
     \end{tablenotes}
\end{threeparttable}
}
   \end{center}

%%%%%%%%%%%%%%%%%%%%%%%%%%%%%%%%%%%%%%%%%%%%%%%%%%%%%%%%%%%%%%%%%%%%%%%%%%%%%%%%%%%%%%%%%%%%%%%% 新表格

\begin{center}

\resizebox{1\textwidth}{!}{

\begin{threeparttable}
  \small
  \caption{Estimation of fuel demand Elasticities}
  \centering % centering table

 \label{tab:6}

 \begin{tabular}{cccccccc}

 \hline

\multicolumn{1}{c}{\multirow{2}{*}{Vehicle Types}} &  \multicolumn{2}{c}{Fuel demand elasticities based AIDS model\tnote{a}} & \multicolumn{2}{c}{Fuel demand elasticities based double-log model \tnote{b}}  \\
\cmidrule(l){2-3} \cmidrule(l){4-5} %\cmidrule(l){6-8}
                 &  Own-price & Expenditure & Own-price &  Expenditure  \\
  \hline
   LPV-D  & $-1.028(0.033)$  & 0.525(0.022)  & $-0.621(0.012)$  & $0.456(0.044)$   \\
   MPV-G  & $-1.003(0.018)$  & 0.725(0.013)  & $-1.779(0.034)$  & $1.145(0.036)$   \\
   SPV-G  & $-0.552(0.003)$  & 1.676(0.036)  & $-0.871(0.103)$  & $1.196(0.026)$  \\
   MNPV-G & $-0.629(0.020)$  & 0.466(0.032)  & $-1.738(0.040)$  & $0.697(0.052)$  \\
   HDT-D  & $-0.823(0.065)$  & 0.561(0.059)  & $-0.968(0.192)$  & $0.601(0.071)$   \\
   MDT-D  & $-0.507(0.036)$  & 0.659(0.016)  & $\emph{-0.326(0.201)}$  & $0.358(0.074)$ \\
   LDT-D  & $-0.799(0.055)$  & 1.103(0.012)  & $-0.471(0.198)$  & $0.561(0.073)$   \\
   MNT-G  & $-0.657(0.057)$  & 0.512(0.030)  & $-0.734(0.051)$  & $0.637(0.081)$   \\
   PB-D   & $-0.442(0.024)$  & 0.209(0.038)  & $-0.417(0.034)$  & $0.382(0.034)$   \\
   Taxi-G & $-1.263(0.064)$  & 0.579(0.011)  & $-2.365(0.146)$  & $0.872(0.038)$  \\
  \hline
 \end{tabular}
\footnotesize

 \begin{tablenotes}
   \item[a] The results are estimated by the AIDS model. Values in parentheses represent standard errors. All elasticities are statistically significant at the 5\% level.
   \item[b] The results are estimated by the double-log model. Values in parentheses represent standard errors. The own-price elasticity value in italic (i.e. fuel demand own-price elasticity of MDT-D) is not statistically significant at the 10\% level. All other elasticities are statistically significant at the 5\% level.
\end{tablenotes}

\end{threeparttable}
}
\end{center}

%\vspace{12 pt}
Given the importance of fuel to the nation's economy, especially for China, the fuel demand elasticities tend to be smaller in the short term at the national level. Even for some studies which only estimate the transport sector fuel demand elasticity, different types of vehicles have not been studied separately. Taking the taxi markets in China as an example, because of restrictions on entry to the local markets by the taxi operator licenses, the taxi markets tend to be monopolized by several local taxi companies, which usually indirectly contracted to individuals who are responsible for daily operations\footnote{ Usually in the form of indirect contracting: Taxi companies are responsible for formalities and the acquisition of taxis, and individual contractors are responsible for the daily operations. Contractor on schedule to pay a taxi company contracted fee what is commonly known as ``\emph{Fen-zi-qian}''.}. Increases in gasoline price will directly lead to rising taxi costs. The taxi contractors will accordingly reduce the non-transport rates to improve operational efficiency. Some of the contractors may even withdraw from taxi contract, which results in reducing of gasoline consumption. Additionally, the rising gasoline costs will raise taxi fares (e.g., taxi rates per mileage and fuel surcharges rise), leading to a decrease in the demand for taxi service. Conversely, when the gasoline price drops, taxi drivers are more inclined to take the initiative to look for passengers, which may increase fuel consumption. Reduction of taxi fare also helps to attract passengers. As a consequence, gasoline demand from taxis is more sensitive to price changes than those from other vehicle types are.

\subsection{Pollution emissions elasticities}

From Eqs.(\ref{pi}) and (\ref{eta}), air pollution emissions elasticities (please see Table \ref{tab:61}) are calculated by using the estimated fuel demand elasticities\footnote{For calculation of air pollution emissions elasticities, this paper only uses fuel demand elasticities by the AIDS model.} as well as pollution emission factors (see Table \ref{tab:4}). The pollution emissions price elasticities indicate the changes of air pollution emissions ($\mathrm{CO}$, $\mathrm{NO}_{x}$, and $\mathrm{PM}_{2.5}$) in response to variations in fuel prices for the studied vehicles groups, while the pollution emissions expenditure elasticities reflect the combined effect of all vehicles expenditure elasticities in the fuel demand system. Because this study focuses on the pollution emissions and related health impacts caused by fuel price changes, more emphases are placed here on the pollution emissions price elasticities.

Similar to the price and expenditure elasticities of fuel demand, price elasticities of air pollution emissions are all negative, while expenditure elasticities are positive. This suggests that, if fuel (gasoline and diesel) prices rise, the road transport sector will reduce fuel demand and consumption, which cause reduction of pollution emissions. Similarly, with all vehicles fuel expenditure increasing, the pollution emissions will increase. Additionally, the price elasticities of pollution emissions (in absolute values) are much lower than corresponding price elasticities of fuel demands reported in Table \ref{tab:6}. A plausible explanation is that the  emissions of specific air pollutant might be jointly influenced by all types of vehicles.

%%%%%%%%%%%%%%%%%%%%%%%%%%%%%%%%%%%%%%%%%%%%%%%%%%%%%%%%%%%%%%%%%%%%%%%%%%%%%%%%%%%%%%%%%%%%%%%

\begin{center}

%\resizebox{1\textwidth}{!}{

\begin{threeparttable}
  %\small
  \caption{Estimation of pollution emissions elasticities}
  \centering % centering table

 \label{tab:61}

 \begin{tabular}{cccc}

 \hline

    Vehicle Types  &    $\mathrm{CO}$ &  $\mathrm{NO}_{x}$ & $\mathrm{PM}_{2.5}$ \\

  \hline
     &   \multicolumn{3}{c}{\multirow{1}{*}{Price elasticities}}    \\
   LPV-D   & $-0.007$ & $-0.015$ & $-0.005$ \\
   MPV-G   & $-0.025$ & $-0.004$ & $-0.014$ \\
   SPV-G   & $-0.140$  & $-0.041$ & $-0.183$ \\
   MNPV-G  & $-0.117$  & $-0.011$ & $-0.011$ \\
   HDT-D   & $-0.018$  & $-0.350$ & $-0.117$  \\
   MDT-D   & $-0.010$  & $-0.224$ & $-0.013$  \\
   LDT-D   & $-0.066$  & $-0.088$ & $-0.136$ \\
   MNT-G   & $-0.004$  & $-0.001$ & $-0.004$  \\
   PB-D    & $-0.018$  & $-0.297$ & $-0.016$ \\
   Taxi-G  & $-0.026$  & $-0.005$ & $-0.057$ \\
     \hline
       % \cline{2-4}
        & \multicolumn{3}{c}{\multirow{1}{*}{Expenditure elasticities}}    \\
           & $0.958$  & $0.705$ & $1.038$  \\
  \hline
 \end{tabular}
\footnotesize
 \begin{tablenotes}
    \item[]  For estimation of air pollution emissions elasticities, this paper only uses fuel demand elasticities by the AIDS model.
\end{tablenotes}

\end{threeparttable}
%}
\end{center}

\vspace{10 pt}

%%%%%%%%%%%%%%%%%%%%%%%%%%%%%%%%%%%%%%%%%%%%%%%%%%%%%%%%%%%%%%%%%%

Specifically, the $\mathrm{CO}$ emission price elasticities (in absolute values) of SPV-G and MNPV-G are $-0.140$ and $-0.117$, respectively, which are much higher than other vehicles. The results suggest that $\mathrm{CO}$ emissions from these vehicle types are more sensitive to gasoline prices than the other types are. Moreover, the $\mathrm{NO}_{x}$ emission price elasticities of HDT-D ($-0.350$), MDT-D ($-0.224$) and PB-D ($-0.297$) are larger than those of other vehicle types. Especially, $\mathrm{NO}_{x}$ emission price elasticities of the diesel vehicles (i.e., HDT-D, MDT-D and PB-D) are significantly greater than those of gasoline vehicles, implying that, compared to gasoline, diesel price has larger impact on $\mathrm{NO}_{x}$ emissions from vehicles. While, for the $\mathrm{PM}_{2.5}$ emission price elasticities, the values of SPV-G ($-0.183$), HDT-D ($-0.117$) and LDT-D ($-0.136$) are greater. To some extent, these findings are consistent with the reality of China's vehicle emissions. In China, $\mathrm{NO}_{x}$ and $\mathrm{PM}$ emissions from trucks (diesel vehicles) are generally higher than those from passenger cars (gasoline vehicles), while the $\mathrm{CO}$ emissions from passenger cars (gasoline vehicles) are greater than those from trucks (diesel vehicles). Particularly, heavy trucks are the major contributors to $\mathrm{NO}_{x}$\footnote{Source: China Vehicle Emission Control Annual Report 2013, released by the Ministry of Environmental Protection of China, http://www.zhb.gov.cn/gkml/hbb/qt/201401/t20140126\_266973.htm}.

From Eq.(\ref{pi}), the emissions price elasticities depend on the fuel demand price elasticities and proportion of pollution emissions associated with vehicle population, VMT, emission factors, etc\footnote{Some other factors may also indirectly affect vehicle emissions, such as, in particular, vehicle emissions-control technologies, and emissions standards \citep{Pandian2009}.}. For instance, as shown in Table \ref{tab:4}, pollution emission factors of PB-D, HDT-D, and LDT-D are relatively large. While, PB-D and HDT-D have the larger VMTs (see Table \ref{tab:4}), and the SPV-G has the largest vehicle population in China.

\subsection{Air pollution emissions and health losses from the road transport sector}

In this section, the air pollution emissions and residents' health losses\footnote{In this study, the residents' health losses include the premature deaths cases caused by acute exposure, and corresponding economic losses.} are estimated under different scenarios based on real-world fuel price changes. To be more specific, firstly, we estimate the changes of air pollution emissions and corresponding concentrations by Eq.(\ref{qtq1}) under different scenarios extracted from real world oil price fluctuations (see Panel A of Table \ref{tab:8}) based on elasticities of fuel demand and air pollution emissions (Table \ref{tab:6} and Table \ref{tab:61}). Secondly, from the health effect models (for linear model, please see Eq.(\ref{AM}); for non-linear model, please see Eqs.(\ref{AMrr}), (\ref{AF}) and (\ref{AMe}))\footnote{In order to avoid the (possible) limitations of the linear health effect model, and to obtain more robust and accurate health loss estimates, we also apply the non-linear models (i.e. Eqs.(\ref{AMrr}), (\ref{AF}) and (\ref{AMe})) to evaluate the health losses, and then compare the results obtained from different methods (please see the Panel B of Table \ref{tab:8}).}, we estimate the residents' premature death cases caused by air pollution emissions from road transport sector. Finally, by using the Eq.(\ref{VOSL}), the corresponding residents' health economic losses are evaluated (please see the Panel B of Table \ref{tab:8}).

As mentioned in Section 3, in China, gasoline and diesel prices are mainly affected by international oil, especially the international crude oil prices (e.g. Brent and WTI), whose fluctuations are transmitted to the prices of downstream oil products. Particularly, after the 2008 financial crisis, crude oil prices experienced drastic fluctuations. Therefore, based on the real-world international oil price fluctuations, we introduce four scenarios from the real-world price shocks to fuel prices. In Scenarios 1 \& 3, prices of gasoline and diesel simultaneously increase by 25\% (the extreme positive price shock scenario) and 5\% (the average positive price shock scenario), respectively. In Scenarios 2 \& 4, prices of gasoline and diesel simultaneously declined by 35\% (the extreme negative price shock scenario) and 3\% (the average negative price shock scenario), respectively (see Table \ref{tab:8}).

Results indicate that, when gasoline and diesel prices increase (Scenarios 1 \& 3), air pollution emissions (emissions quantities and concentrations) and residents' health losses caused by road transport decline correspondingly. In contrast, when gasoline and diesel prices decline (Scenarios 2 \& 4), the air pollution emissions and corresponding losses increase. To be more specific, compared with the base year 2012, when gasoline and diesel prices rise 25\% simultaneously (Scenario 1), emissions of $\mathrm{CO}$, $\mathrm{NO}_{x}$, and $\mathrm{PM}_{2.5}$ from road transport decrease by 372690, 166020, and 5620 tons, respectively. Furthermore, the total reduction of air pollution emissions from the sector reached 544330 tons. The concentrations drop by 0.003 $m g/m^{3}$ , 0.960 $\mu g/m^{3}$, and 0.079 $\mu g/m^{3}$, respectively. The results of residents' health losses from the linear health effect model (see Eq.(\ref{AM})) show that reductions of residents' premature deaths caused by acute exposure are 1154, 12080 and 322 cases, and that the corresponding economic losses decrease about 987.292, 10336.235 and 275.649 million RMB yuan, respectively. As a result, the number of premature deaths decrease by 13556; and economic losses are reduced 11599.176 million yuan caused by air pollution emissions from the sector. As a comparison, the results from the non-linear health effect models (see Eqs.(\ref{AMrr}), (\ref{AF}) and (\ref{AMe})) show that the reductions of residents' premature deaths are 1141, 11520 and 321 cases, and that the corresponding economic losses decrease about 976.452, 9856.978 and 274.995 million yuan; the premature deaths decrease by 12982 cases and economic losses by 11108.4 million RMB yuan. For more details, please see Table \ref{tab:8}. It is obvious that the results based on different health effect models are qualitatively consistent without significant difference, which further indicates that results of this paper are reliable and robust.

%%%%%%%%%%%%%%%%%%%%%%%%%%%%%%%%%%%%%%%%%%%%%%%%%%%%%%%%% 新计算结果%%%%%%%%%%%%%%%%%%%%%%%%%%%%%%%%%%%%%%%%%%%%%%%%

\begin{sidewaystable}

\resizebox{1\textwidth}{!}{

\begin{threeparttable}
\small

  \caption{Air pollution emissions and residents' health losses from the road transport sector\tnote{a}}
  \centering % centering table
 \label{tab:8}       % Give a unique label

\begin{tabular}{ccccccccc}
  \hline

 \multicolumn{9}{l}{\multirow{1}{*}{Panel A: Estimation of air pollution emissions }}\\
   \multicolumn{2}{c}{\multirow{2}{*}{Scenarios\tnote{b}}}  & \multicolumn{2}{c}{\multirow{1}{*}{$\mathrm{CO}$}}& \multicolumn{2}{c}{\multirow{1}{*}{$\mathrm{NO}_{x}$}} & \multicolumn{2}{c}{\multirow{1}{*}{$\mathrm{PM}_{2.5}$}} & \multicolumn{1}{c}{\multirow{2}{*}{\textbf{Total}\tnote{e}}} \\
   \cmidrule(l){3-4} \cmidrule(l){5-6} \cmidrule(l){7-8}
    &    & $\triangle$Quantity (\%)\tnote{c} & $\triangle$Concentration(\%)\tnote{d}  & $\triangle$Quantity (\%)\tnote{c}  & $\triangle$Concentration(\%)\tnote{d} & $\triangle$Quantity (\%)\tnote{c} & $\triangle$Concentration(\%)\tnote{d} & \\
    \hline
  \multicolumn{2}{c}{\multirow{1}{*}{Scenario 1($+25\%$)}} & $-37.269(-1.074)$ & $-0.003(-0.248)$ & $-16.602(-2.594)$  & $-0.960(-2.042)$ & $-0.562(-1.390)$  & $-0.079(-0.177)$  & $-54.433$\\
  \multicolumn{2}{c}{\multirow{1}{*}{Scenario 2($-35\%$)}}& $52.177(1.503)$ & $0.005(0.347)$ & $23.243(3.632)$ & $1.344(2.859)$ & $0.786(1.946)$  & $0.111(0.248)$  & $76.206$ \\
  \multicolumn{2}{c}{\multirow{1}{*}{Scenario 3($+5\%$)}} & $-7.454(-0.215)$ & $-0.001(-0.050)$ & $-3.320(-0.519)$ & $-0.192(-0.408)$ & $-0.112(-0.278)$  & $-0.016(-0.035)$  & $-10.887$\\
  \multicolumn{2}{c}{\multirow{1}{*}{Scenario 4($-3\%$) }}& $4.472(0.129)$   & $0.001(0.030)$   & $1.992(0.311)$    & $0.115(0.245)$   & $0.067(0.167)$    & $0.010(0.021)$    &  $6.532$\\

  \hline
\multicolumn{9}{l}{\multirow{1}{*}{Panel B: Evaluation of residents' health losses}} \\
 \multicolumn{2}{c}{\multirow{2}{*}{Scenarios\tnote{f}}}  & \multicolumn{2}{c}{\multirow{1}{*}{$\mathrm{CO}$}}& \multicolumn{2}{c}{\multirow{1}{*}{$\mathrm{NO}_{x}$}} & \multicolumn{2}{c}{\multirow{1}{*}{$\mathrm{PM}_{2.5}$}} & \multicolumn{1}{c}{\multirow{2}{*}{\textbf{Total}\tnote{i}}} \\
 \cmidrule(l){3-4} \cmidrule(l){5-6} \cmidrule(l){7-8}
     &  & $\triangle$Premature deaths\tnote{g} & $\triangle$ losses\tnote{h}  & $\triangle$Premature deaths\tnote{g} & $\triangle$ losses\tnote{h} & $\triangle$Premature deaths\tnote{g} & $\triangle$ losses\tnote{h}  &\\

\hline
  Scenario 1($+25\%$) & linear case& $-1154$  & $-987.292$  & $-12080$  & $-10336.235$  & $-322$ & $-275.649$ & $-13556(-11599.176)$\\
                              &  non-linear  case& $-1141$  & $-976.452$  & $-11520$  & $-9856.978$  & $-321$ & $-274.995$ & $-12982(-11108.4)$\\

  Scenario 2($-35\%$)& linear  case& $1615$ & $1382.209$  & $16912$  & $14470.729$   & $451$ & $385.909$ & $18979(16238.847)$\\
                 & non-linear  case& $1597$ & $1366.837$  & $16104$  & $13779.1231$  & $450$ & $384.907$ & $18151(15530.94)$\\
 Scenario 3($+5\%$)& linear   case& $-231$ & $-197.458$  & $-2416$  & $-2067.247$   & $-64$ & $-55.130$ & $-2711(-2319.835)$\\
                        & non-linear  case& $-228$ & $-195.281$  & $-2303$  & $-1970.412$   & $-63$ & $-54.998$ & $-2594(-2220.69)$\\
   Scenario 4($-3\%$) & linear   case& $138$  & $118.475$   & $1450$   & $1240.348$    & $39$  & $33.078$  & $1627(1391.901)$\\
                & non-linear  case& $137$  & $117.166$   & $1381$   & $1182.011$    & $39$  & $32.999$  & $1557(1332.176)$\\
   % \multicolumn{1}{c}{\multirow{2}{*}{Scenario 4($-3\%$) }} &linear  & $138$  & $118.475$   & $1450$   & $1240.348$    & $39$  & $33.078$  & $1627(1391.901)$\\
     %         & non-linear & $137$  & $117.166$   & $1381$   & $1182.011$    & $39$  & $32.999$  & $1557(1332.176)$\\

  \hline

\end{tabular}

 \begin{tablenotes}
    \item[a] We use the 2012 data as the baseline data.
    \item[b] Scenarios 1 \& 3 respectively refer to scenarios that prices of gasoline and diesel simultaneously rose by 25\% and 5\%. Scenarios 2 \& 4 respectively refer to prices of gasoline and diesel simultaneously declined by 35\% and 3\%.
   \item[c] $\triangle$Quantity is the changes of pollution emissions from road transport (unit: 10,000 tons). The values in the parentheses stand for percentage changes.
   \item[d] $\triangle$Concentration is the changes of air pollutant concentrations (unit: $m g/m^{3}$ for $\mathrm{CO}$, $\mu g/m^{3}$ for $\mathrm{NO}_{x}$ and $\mathrm{PM}_{2.5}$). The values in the parentheses stand for percentage changes.
   \item[e] The total amount of changes in pollution emissions from the road transport sector (unit: 10,000 tons).
    \item[f] Scenarios 1 \& 3 respectively refer to scenarios that prices of gasoline and diesel simultaneously rose by 25\% and 5\%. Scenarios 2 \& 4 respectively refer to prices of gasoline and diesel simultaneously declined by 35\% and 3\%. The linear case means estimated premature deaths based on linear health effects model (\ref{AM}) and corresponding economic losses. The non-linear case means estimated premature deaths based on non-linear health effects model (\ref{AMe}) and corresponding economic losses.
    \item[g] The premature deaths cases caused by acute exposure.
   \item[h] Unit: million of RMB yuan.
    \item[i] The total number of changes in residents' premature deaths and corresponding economic losses (in the parentheses, unit: million of RMB yuan) caused by air pollution from the road transport sector. It is very complicated about the health damages caused by the joint effects of different air pollutants, on which there is no consensus and solid methods to deal with this issue; for simplicity, these effects are ignored in this study.

    \end{tablenotes}

\end{threeparttable}
}
\end{sidewaystable}

\section{Conclusions and Future work}

In recent years, China's rapid economic development resulted in serious air pollution. In particular, the road transport sector has become one of the major emitters, which caused substantial losses to residents' health. In the broad context of the drastically fluctuating international oil prices and the Chinese economic transition, it is extremely important to study the transmission mechanisms between the oil price changes and residents' health losses in the transitional China. However, by now, there is no previous studies to address this critically important issue in existing literature. Therefore, inspired by \cite{Blanchard2001} and \cite{Li2005}, we propose a key assumption that China's provincial economies are independent ``economic entities''. In this foresaid context, we estimate the China's road transport fuel demand system, and investigate the impacts of pollution emissions and residents' health losses from road transport under different scenarios extracted from real world price shocks (Scenarios 1-4). Our main findings can be summarized as follows:

First of all, we investigate the China's road transport fuel demand system by estimating the price elasticities of fuel demand from different vehicle categories. The results indicate that all the own-price elasticities are negative and statistically significant, and vary across different vehicle types. The expenditure elasticities are positive and statistically significant. The own-price elasticities (in absolute values) for small passenger vehicles (G), mini passenger vehicles (G), heavy duty trucks (D), medium duty trucks (D), light duty trucks (D), mini trucks (G) and public buses (D) are less than 1, which means that these are not sensitive to fuel price changes. However, own-price elasticities (in absolute values) for large passenger vehicles (D), medium passenger vehicles (G) and Taxi (G) are all greater than (or equal to) 1, implying that these are price-sensitive.

Secondly, the pollution emissions price elasticities indicate the changes of air pollution emissions ($\mathrm{CO}$, $\mathrm{NO}_{x}$, and $\mathrm{PM}_{2.5}$) in response to the variations in fuel prices. The price elasticities of air pollution emissions are all negative, while pollution emission expenditure elasticities are positive. Specifically, for $\mathrm{CO}$, the emissions price elasticities of small passenger vehicles (G) and mini passenger vehicles (G) are comparatively larger than those of other vehicle types. For $\mathrm{PM}_{2.5}$, the emissions price elasticities of small passenger vehicles (G), heavy duty trucks (D) and light duty trucks (D) are comparatively larger.  $\mathrm{NO}_{x}$ emission price elasticities of diesel vehicles are significantly greater than those of gasoline vehicles, implying that $\mathrm{NO}_{x}$ emissions are more sensitive to changes of diesel price than that of gasoline price.

Finally, in the context of drastic oil price changes, by using four different health effect models (i.e. one linear model and three non-linear models), this study estimates the air pollutant emissions and residents' health losses under different price change scenarios. The results from the linear and non-linear models all indicate that the increases of gasoline and diesel prices cause the declines in the road transport fuel demand, and consequently in residents' health losses, while the declines in gasoline and diesel prices result in the increases in road transport fuel demand, the corresponding health losses. By comparing the results from different (linear and nonlinear) models, we found  that the results are reliable and robust.

In summary, from the perspective of road transport pollution emissions, this paper is the first attempt to provide a framework for understanding the underlying mechanisms how the oil price changes are transmitted to the demand system, and thereby cause residents' health losses (or gains) in the transitional China.

\newpage
\section*{Appendix: China's Fiscal decentralization, road transport sector, and  ``independent economic entities" hypothesis}

The implementation of fiscal decentralization reforms are widely regarded as one of the most important driving forces behind China's growth ``miracle'' \citep{Blanchard2001, Feltenstein2005, Lin2000, Jin2005}. Although China's fiscal system consists of many hierarchical levels of government: central, provincial, prefecture, county, and township, the most important and fundamental hierarchical separation is the central--provincial fiscal decentralization\citep{Jin2005}, which is the key move to more local economic freedom and further sub-level fiscal decentralization.

Generally speaking, China's fiscal decentralization reforms have undergone three main phases: the pre-reform phase prior to 1979, the transitional phase of 1980 --1993, and the post 1994 phase \footnote{Prior to the reform of 1979, the China's fiscal policy was called as ``unified revenue collection and unified spending'' (\emph{Tong-zhi-tong-shou}). In the transitional phase, the ``fiscal contracting system'' (\emph{Cheng-bao-zhi}) was implemented. In practice, China's fiscal decentralization reforms were very complicated and they varied across provinces and over time. For a detailed discussion of China's fiscal decentralization, see \cite{Blanchard2001, Jin2005, Lin2000, Zhang1998}.} \citep{Jin2005}. From 1994, the ``separating tax system''(\emph{Fen-shui-zhi}) was implemented, which is considered as the most important and extensive phase. In the ``separating tax system'', the ``local revenue'' was redefined as revenues from local taxes and local portion of the shared taxes. As a result, provincial governments became independent fiscal entities that had both responsibility for local expenditures and the unprecedented right to use the revenue that they retained \citep{Jin2005,Lin2000}. By decentralizing its formerly centrally planned fiscal system through fiscal decentralization, China not only reduced its central government's own fiscal burden, but also considerably strengthened the fiscal incentives of its provincial governments, and thereby motivated the local governments to promote local economic and infrastructure development \citep{Lin2000, Jin2005}.

After the fiscal decentralization, local (say, provincial) leaders in China are usually assessed mainly based on their economic performance in promoting the local economy (also termed as "promotion tournament") \citep{Blanchard2001, Li2005, Oi1995}. \cite{Blanchard2001} find that local governments in China have actively contributed to the growth of new firms, which can partly explain the highest growth rate of China's GDP. \cite{Li2005} believe that China is run in such a way that its provincial leaders behave like line managers of a large-scale multi-level company and ``their internal career mobility is closely tied to their economic performance". Similarly, \cite{Oi1995} even concludes that akin to a large multi-level corporation, Chinese central government per se acts as the headquarters of a corporate, while the provincial authorities behave as the regional headquarters. According to their theories, local officials in China are in fact a board of directors and sometimes more directly as the chief executive officers; each successive level of governments, consequently, is fiscally independent and thus expected to maximize its economic performance \citep{Li2005, Oi1995}.

Fiscal decentralization stimulates the local governments' willingness to promote economic development, especially by large-scale investment in road transport sector. According to inter-jurisdictional competition theory \citep{Head1996, Tiebout1956, Keen1997, Wilson1986} and common-pool problem \citep{Stein1999}, the local governments prefer to increase expenditure on infrastructure to obtain regional competitive advantage and thereby to promote local economic growth. In China, such shift of fiscal power and responsibility to lower levels of government increases economic efficiency, because local governments have informational advantages over the central government concerning resource allocation \citep{Lin2000}, encouraging local expenditures heavily on infrastructure to attract mobile capital and boost local economies \citep{Jia2014}.

The growths of the Chinese central and local governments fiscal revenue and expenditures were both very slow for a long time. However, after the implementation of the ``separating tax system'' in 1994, the growths significantly accelerated. For example, in 1978, the fiscal revenues were only 17.577 billion RMB yuan for the central government, and 95.649 billion RMB yuan for local governments; by contrast, in 1994, central and local governments fiscal revenues reached 290.65 and 231.16 billion RMB yuan, respectively. In 2012, the fiscal revenues increased dramatically to be 6107.829 billion RMB yuan for local governments, and 5617.523 billion RMB yuan for the central government. Meanwhile, the local fiscal expenditures exceed dominantly the central ones after 1994. For example, in 2012, the former reached 10718.83 billion RMB yuan; while as a comparison, the latter was only 1876.463 billion RMB yuan.

More importantly, the central and local governments continue to increase investment in road transport infrastructure after the year of 1994. In particular, the investments from local governments, again, exceed dominantly those from the central government. For instance, in 2011, the investments  in road and public transports from local governments were 1337.671 billion and 218.318 billion RMB yuan, while those from the central government were 47.964 billion and 4.211 billion RMB yuan respectively\footnote{Data source: NBSC, http://data.stats.gov.cn/index}. As a result, the total road and highway mileages of China witnessed rapid growths, and increased from 1.12 million and 1,600 kilometers (in 1994) to 4.11 million and 84,946 kilometers (in 2011), respectively.

Furthermore, with the fiscal decentralization and increasing investment on infrastructure in road transport sector, China's on-road vehicle population keeps a rapid increase. The average yearly growth of vehicle population after the ``separating tax system'' (post-1994 phase) was approximately 14.78\%\footnote{The average annual growth is calculated basing on annual civil vehicles population from 1994 to 2013.} from 1994 to 2013. More specifically, the stocks of civil vehicles in China were 9.42 million in 1994 and 126.70 million in 2013, which were almost 70 and 13 times more than those in 1980 and 1994, respectively. The growths of vehicle stocks in many regions of China were in a similar situation. For example, the vehicle population in Beijing keeps increasing at an average annual rate of 14.4\% since 1990, and has been now more than 5 million\footnote{Source: Beijing Statistical Yearbook 1993-2012.}. Along with the significant increase of road infrastructure and motor vehicle stock, China's freight and passenger traffic volumes increase rapidly, which has gradually become the dominant part of the transport systems in China. Specifically, the passenger traffic volume alone accounted for 55\% out of the whole transport system in 2012, while the freight traffic volume made up 13\% in 2000 to 35\% in road transport in 2012.

As aforesaid, China is facing severe pressures on the environment deterioration and health damages attributed to the activities of the road transport sector. Therefore, as the ``visible hand'', more and more provincial governments in China began to implement strict transport policies such as ``plate traffic restriction''\footnote{This policy imposes a restriction on vehicle usage that vehicles with license plates ending with even numbers can operate on roads only on even-numbered dates, and odd numbers on odd-numbered dates. For more detailed regulations, see the Beijing Traffic Management Bureau, http://www.bjjtgl.gov.cn/publish/portal0/}, ``vehicle purchase limit''\footnote{This policy puts a limit on the quantity of new motor vehicles, and force eligible car buyers, who already meet strict requirements, to apply for car license by network lot drawing. For more details and other regulatory policies imposed on car ownership, see the Beijing Traffic Management Bureau, http://www.bjjtgl.gov.cn/publish/portal0/} and ``vehicle emission standards''. To cite an example, Beijing implemented the ``plate traffic restriction'' in 2009, ``vehicle purchase limit'' in 2011, and introduced Euro-I emission standards in 1999, Euro-II standards in 2003, and Euro-III standards in 2006. Many other provincial governments (e.g., Shanghai and Tianjin) imposed (or are going to implement) similar policies to restrict the vehicle usage, vehicle ownership and compulsory vehicle emission standards. In brief, after fiscal decentralization, local governments have very important impacts on the road transport sector by various means, such as investing road infrastructures, encouraging or restricting vehicle usage or ownership, and imposing tight traffic regulations and laws, etc.

In summary, China, as the world's largest transition economy, whose economic development has its uniqueness. Especially, the fiscal decentralization considerably strengthened the fiscal incentives of provincial governments, and provided economic stimuli for these local governments to promote local economies. As a result, the former centrally controlled planned economy transformed gradually into a socialist market economy consisting of many regional planned economies controlled by sub-national governments \citep{Jin2005a}, which exhibit many characteristics of a business corporation, with local officials acting as the equivalent of a board of directors \citep{Oi1995}. Therefore, the role of local governments in China is the key to understand the Chinese conundrum of its rapid economic development and residents' health losses caused by environmental deterioration. Unfortunately, the current studies in existing literatures failed to capture the essence due to the ignorance of the role of local government as an economic entity, and thereby too often arrived at dubious (sometimes even misleading) conclusions and/or policy suggestions. Therefore, we propose a key assumption that China's provincial economies are independent ``economic entities'' acting as the equivalent of a business company in the context of China's fiscal decentralization.

\newpage
\section*{References}

\bibliographystyle{model5-names}
\bibliography{myref2015cer}

\end{document}